\documentclass[11pt,draftclsnofoot,onecolumn]{IEEEtran}
\usepackage{enumerate}
\usepackage{eurosym}
\usepackage{amsmath}
\usepackage{amsfonts}
\usepackage{amssymb}
\usepackage{graphicx}
\usepackage{epstopdf}
\usepackage{setspace}
\usepackage{arydshln}
\usepackage{float}
\usepackage[ruled,linesnumbered]{algorithm2e}
\usepackage[noend]{algpseudocode}
\usepackage[english]{babel}
\usepackage{amsthm}
\usepackage{enumitem}
\usepackage{xcolor}
\usepackage{soul}
\usepackage{url}
\usepackage{booktabs}

\makeatletter
\renewcommand{\fnum@figure}{Fig. \thefigure}
\makeatother

\makeatletter
\newcommand{\removelatexerror}{\let\@latex@error\@gobble}
\makeatother

\begin{document}
	\title{Fast Data-Driven Adaptation of Radar Detection via Meta-Learning} 
	\author{\IEEEauthorblockN{Wei Jiang\IEEEauthorrefmark{1}, Alexander M. Haimovich\IEEEauthorrefmark{1}, Mark Govoni\IEEEauthorrefmark{2}, Timothy Garner\IEEEauthorrefmark{2}, and Osvaldo Simeone\IEEEauthorrefmark{3}}\\ \IEEEauthorblockA{\IEEEauthorrefmark{1}CWiP, New Jersey Institute of Technology, Newark, New Jersey 07102, USA \\ \IEEEauthorrefmark{2} U.S. Army DEVCOM Army Research Laboratory, Adelphi and Aberdeen Proving Ground, Maryland  20783 and 21005, USA\\ \IEEEauthorrefmark{3}KCLIP, Department of Engineering, King's College London, London, WC2R 2LS, UK} }
	\maketitle
	
	\thispagestyle{plain}
	\pagestyle{plain} 

	\begin{abstract}
		This paper addresses the problem of fast learning of radar detectors with a limited amount of training data. 
		In current data-driven approaches for radar detection, re-training is generally required when the operating environment changes, incurring large overhead in terms of data collection and training time. 
		In contrast, this paper proposes two novel deep learning-based approaches that enable fast adaptation of detectors based on few data samples from a new environment. 
		The proposed methods integrate prior knowledge regarding previously encountered radar operating environments in two different ways. 
		One approach is based on transfer learning: it first pre-trains a detector such that it works well on data collected in previously observed environments, and then it adapts the pre-trained detector to the specific current environment. 
		The other approach targets explicitly few-shot training via meta-learning: based on data from previous environments, it finds a common initialization that enables fast adaptation to a new environment. 
		Numerical results validate the benefits of the proposed two approaches compared with the conventional method based on training with no prior knowledge. 
		Furthermore, the meta-learning-based detector outperforms the transfer learning-based detector when the clutter is Gaussian.	
	\end{abstract}

	\section{Introduction}
	Radar detector design has been a problem of long-standing interest in the radar literature \cite{Meyer1973}. 
	Optimal target detection is achieved by a test derived from the Newman-Pearson (NP) criterion, which guarantees the highest probability of detection for a given probability of false alarm \cite{Kay1998}. For example, it is well known that the square law detector is optimal in the NP sense when targets, clutter and interference follow Gaussian distributions \cite{Richards2005}. However, such Gaussian models do not always reflect the actual operating conditions. For example, in high-resolution radar applications and/or at low grazing angles, the probability of observing large values of the clutter amplitude is greater than based on Rayleigh statistics \cite{Richards2010}. Accordingly, the clutter amplitude is commonly modeled by non-Gaussian distributions, such as Weibull, K, and lognormal distributions \cite{Gini2007}. In most cases involving heavy-tailed non-Gaussian models, the structure of optimal detectors requires intractable numerical integrations, which makes the implementation of such detectors computationally intensive \cite{Farina1986}. Moreover, detectors designed based on specific models suffer performance degradation when the actual signals behave differently than their assumed mathematical models \cite{Gini1998}.

	Deep learning has been successfully applied in a variety of fields to solve problems for which reliable mathematical models are unavailable or too complex to yield feasible optimal solutions \cite{osvaldo2}. In the radar field, deep learning-based approaches have been proposed for implementing NP detectors 
	\cite{Moya2013}. These approaches rely on the assumption that the training and the actual operating environments have similar statistical properties. 
	However, in practice, mismatch between training and testing conditions may result in detection performance loss \cite{Wei2019NN}. To deal with this problem, a straightforward approach is to re-train the detector from scratch based on newly collected data from the current operating environment. However, this approach requires large overhead in terms of data collection and training time. In \cite{Wei2021}, the authors train the detector with a mixture of data from different environments to robustify detection performance with respect to the characteristics of observed data.

	Transfer learning and meta-learning are two different learning paradigms in machine learning that can be used to address different operating conditions. 
	In transfer learning, the goal is to extract knowledge from source tasks, so as to improve training efficiency on a target task \cite{transfer}. 
	In contrast, meta-learning, or ``learning to learn", aims to infer an inductive bias given the data from multiple related tasks, to enable efficient training on a new task within a certain class \cite{meta2020}. The inferred inductive bias can take the form of a learning procedure or a prior over the model parameters \cite{osvaldo3}. 
	Notably, reference \cite{MAML2017} proposes the model-agnostic meta-learning (MAML) algorithm, which optimizes the initialization for the parameters of a neural network to enable fast adaptation on a new task. Recently,
	the two learning paradigms have been applied to solve problems in communication systems to achieve fast adaptation, such as resource allocation in wireless networks \cite{resource}, downlink beamforming optimization \cite{beamform}, demodulating over fading channels \cite{osvaldo_journal}, and decoding for convolutional and turbo codes \cite{Mind}.

	In this work, we aim to design detectors that adapt quickly to the radar operating environment based on few data samples. Unlike the techniques in \cite{Moya2013}, the deep learning-based detector design is separated into two stages, i.e., an offline training stage and an adaptation stage. 
	The goal of the offline training stage is to leverage prior knowledge from multiple radar environments via transfer learning or meta-learning. Being offline, this phase may be implemented with large amounts of data. In the adaptation phase, we refine the training based on few data samples collected from the current radar operating environment.
	Specific contributions of this
	work are: (1) we propose a two-stage learning procedure that enables detectors to adapt quickly to the current radar environment; (2) we develop a deep transfer learning-based algorithm for fast adaptation of  radar detection; and (3) we develop a meta-learning-based algorithm by leveraging MAML for the design of fast adaptation of radar detection.
	
	\section{Problem Formulation}
	Consider a pulse-compression radar system, in which the system seeks to detect the presence of a single target over a clutter field.
	The transmitter emits $K$ modulated chips with deterministic complex amplitudes forming a coded waveform $\mathbf{y}=[y_1,\ldots, y_K]^T$. After chip matched filtering and sampling, the discrete-time $K$-dimensional column received signal, for the range cell under test containing a point target, is given by 
	\begin{equation}
		\mathbf{z} = \alpha\mathbf{y} + \mathbf{c} + \mathbf{n},
	\end{equation}
	where $\alpha$ is the complex target gain; $\mathbf{c}$ is the clutter vector; and $\mathbf{n}$ is the noise vector. 
	Detection of the presence of a target in the range cell under test leads to the following binary hypothesis test
	\begin{equation}
		\left\{ \begin{aligned} &\mathcal{H}_0:{\mathbf{z}}={\mathbf{c}}+{\mathbf{n}} \\
			&\mathcal{H}_1:{\mathbf{z}}=\alpha\mathbf{y}+{\mathbf{c}}+{\mathbf{n}},
		\end{aligned} \right.  \label{eq:binary hypo}
	\end{equation}
	where $\mathcal{H}_0$ and $\mathcal{H}_1$ represent the hypotheses under which the target is absent and present, respectively. 
	
\begin{figure}[H]
		\vspace{-8ex} \hspace{16ex} \includegraphics[width=1.4
		\linewidth]{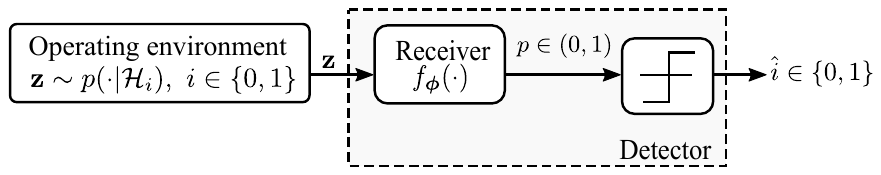} \vspace{-179ex}
		\caption{A deep-learning-based detector. The receiver is implemented as a parametric function $f_{\boldsymbol{\phi}}(\cdot)$ with trainable parameter vector $\boldsymbol{\phi}$.}
		\label{f:detector}
	\end{figure}

	The deep learning-based detection system under study is illustrated in Fig. \ref{f:detector}. The receiver is implemented as a parametric function $f_{\boldsymbol{\phi}}(\cdot)$ with trainable parameter vector $\boldsymbol{\phi}$. The radar operating environment is modeled as a stochastic system that produces the vector $\mathbf{z}\in \mathbb{C}^K$ from a likelihood function $p(\mathbf{z}|\mathcal{H}_i)$. The absence or presence of a target is indicated by the values $i=0$ and $i=1$, respectively. The receiver passes the vector $\mathbf{z}$ through a trainable mapping $p=f_{\boldsymbol{\phi}}(\mathbf{z})$, which produces the scalar $p\in (0,1)$. The final decision $\hat{i}\in \{0,1\}$ is made by comparing the output
	of the receiver $p$ to a hard threshold in the interval $(0,1)$.
	
	In the following, we detail an implementation of the receiver $f_{\boldsymbol{\phi}}(\cdot)$ based on a feedforward neural network. Denote the number of neurons at the input and output layers $M_0$ and $M_L$, respectively.
	A feedforward neural network is a parametric function that defines a mapping $\mathbf{u}_{L}={f}_{\boldsymbol{\phi}}(\mathbf{u}_0)$ from an input real-valued vector $\mathbf{u}_0\in\mathbb{R}^{M_0}$ to an output real-valued vector $\mathbf{u}_L\in\mathbb{R}^{M_L}$ via $L$ successive layers. At the $l$th layer, the intermediate output is 
	\begin{equation}
		\mathbf{u}_l={f}_{\boldsymbol{\phi}_l} (\mathbf{u}_{l-1})= \sigma(\mathbf{W}_l\mathbf{u}_{l-1}+\mathbf{b}_l),
	\end{equation}
	where $\sigma(\cdot)$ represents an element-wise activation function, and $\boldsymbol{\phi}_l=\{\mathbf{W}_l, \mathbf{b}_l\}$ includes the trainable parameters of the $l$th layer consisting of the weight matrix $\mathbf{W}_l$ and the bias vector $\mathbf{b}_l$. The receiver trainable parameter set contains all parameters of the network, and is denoted $\boldsymbol{\phi}=\{\boldsymbol{\phi}_1,\ldots,\boldsymbol{\phi}_L\}$. The input real-valued vector $\mathbf{u}_0$ of the receiver $f_{\boldsymbol{\phi}}(\cdot)$ is the concatenation of the real and imaginary parts of the received vector $\mathbf{z}$. The last layer of the neural network $f_{\boldsymbol{\phi}}(\cdot)$ is selected as a logistic regression layer. The absence or presence of the target is determined by comparing the output of the neural network $f_{\boldsymbol{\phi}}(\cdot)$ with a threshold selected via the false alarm probability.
	
	\section{Two-stage Design of Fast Adaptive Receiver}
\begin{figure}[H]
		\vspace{-8ex} \hspace{10ex} \includegraphics[width=1.5
		\linewidth]{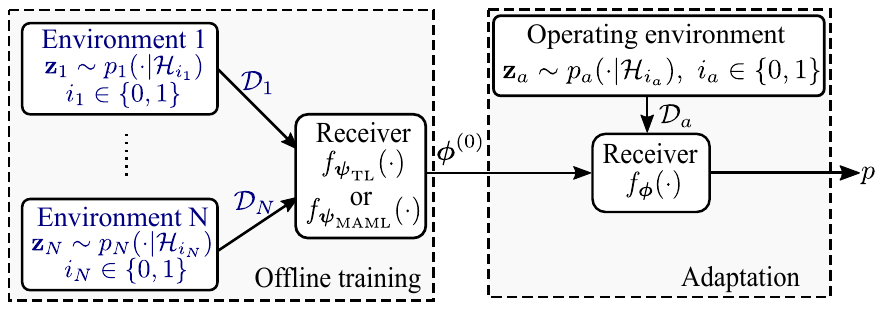} \vspace{-182ex}
		\caption{Illustration of the two-stage design of the fast adaptive receiver. Offline training can leverage either via transfer learning or meta-learning.}
		\label{f:receiver}
	\end{figure}

	This section proposes a two-stage learning procedure that enables the receiver to adapt quickly to the current radar environment. As illustrated in Fig. \ref{f:receiver}, the two-stage learning procedure consists of an offline training stage and an adaptation stage.
	The goal of the offline training stage is to leverage prior knowledge from data collected in multiple radar environments. The offline training stage can be carried out either via transfer learning or meta-learning. The adaptation stage refines the receiver parameter vector based on few samples from the current radar operating environment.  
	
	The offline dataset is a collection of data from $N$ radar operating environments, and is denoted as $\mathcal{D}=\{\mathcal{D}_n \}_{n=1}^N$, where $\mathcal{D}_n=\big\{ \mathbf{z}^{(q)}_n\sim p_n(\mathbf{z}|\mathcal{H}_{i^{(q)}_n}), {i^{(q)}_n}\in\{0,1\}  \big\}_{q=1}^{Q}$ contains $Q$ independent and identically (i.i.d.) training samples collected from the $n$th radar environment. The adaptation dataset is denoted as $\mathcal{D}_a=\big\{ \mathbf{z}^{(q)}_a\sim p_a(\mathbf{z}|\mathcal{H}_{i^{(q)}_a}), {i^{(q)}_a}\in\{0,1\}  \big\}_{q=1}^{Q_a}$, which contains $Q_a$ i.i.d. samples collected from the current radar operating environment. Note that the number of samples used during the adaptation phase could be much smaller than that used during the offline training phase.
	The standard cross-entropy \cite{Moya2013} is adopted as the loss function for the receiver. For any dataset $\mathcal{D}_0=\big\{ \mathbf{z}^{(q)}\sim p(\mathbf{z}|\mathcal{H}_{i^{(q)}}), {i^{(q)}}\in\{0,1\}  \big\}_{q=1}^{Q_0}$ containing  $Q_0$ pairs of received signal $\mathbf{z}$ and target state indicator $i$, the empirical cross-entropy loss is a function of the trainable parameter vector $\boldsymbol{\phi }$, and is given by
	\begin{equation}
		\begin{aligned}
			{\mathcal{L}}_{\mathcal{D}_0}(\boldsymbol{\phi})=\frac{1}{Q_0}\sum_{q=1}^{Q_0} -i^{(q)}\ln f_{\boldsymbol{\phi}}(\mathbf{z}^{(q)})-(1-i^{(q)})\ln\big[1- f_{\boldsymbol{\phi}}(\mathbf{z}^{(q)})\big].
		\end{aligned} \label{eq: rx loss grad}
	\end{equation}

	\subsection{Design of Fast Adaptive Receiver via Transfer Learning}
	The goal of transfer learning is to extract prior knowledge from the offline dataset $\mathcal{D}$ during the offline training stage, so as to improve training efficiency based on the adaptation dataset $\mathcal{D}_a$. In the offline training stage, we aim to find a shared parameter vector, denoted as $\boldsymbol{\psi}_{\text{TL}}$, that performs well over $N$ operating environments. Based on the offline dataset $\mathcal{D}$, the shared parameter vector $\boldsymbol{\psi}_{\text{TL}}$ is obtained by minimizing the sum of empirical loss (\ref{eq: rx loss grad}) over $N$ operating environments. The shared parameter vector $\boldsymbol{\psi}_{\text{TL}}$ is optimized iteratively according to stochastic gradient descent (SGD) rule
	\begin{equation}
		\boldsymbol{\psi}_{\text{TL}} \leftarrow \boldsymbol{\psi} _{\text{TL}}- \beta\nabla_{\boldsymbol{\psi}_{\text{TL}}}\sum_{n=1}^{N} {\mathcal{L}}_{\mathcal{D}_{n}}(\boldsymbol{\psi}_{\text{TL}}), \label{eq: transfer sgd}
	\end{equation}
	where $\beta>0$ is the learning rate.
	
	In the adaptation stage, we refine the training based on the adaptation dataset $\mathcal{D}_a$. The receiver parameter $\boldsymbol{\phi}$
	is updated as follows:
	\begin{equation}
		\boldsymbol{\phi}^{(m)}=\boldsymbol{\phi}^{(m-1)}
		-\alpha {\nabla}_{\boldsymbol{\phi}}{\mathcal{L}}_{\mathcal{D}_a}(\boldsymbol{\phi}^{(m-1)})
		\label{eq: rx sgd}
	\end{equation}
	across iterations $m=1,2,\ldots$ with ${\boldsymbol{\phi}}^{(0)}=\boldsymbol{\psi}_{\text{TL}}$, where $\alpha>0$ is the learning rate. The algorithm of fast adaptive detector via transfer learning is summarized in Algorithm 1.
	
	It is finally noted that the approach described in this subsection is also known as joint learning or fine-tuning (e.g., \cite{osvaldo_journal}). A more general implementation of transfer learning would also assume the availability of data from the operating environment, which is not considered here.
	\begin{algorithm}[]
		\DontPrintSemicolon
		\SetAlgoLined
		\tcc{offline training stage}
		initialize shared parameter vector $\boldsymbol{\psi}_{\text{TL}}$\;
		\While{stopping criterion not satisfied}{
			evaluate the overall empirical loss $\sum_{n=1}^{N} {\mathcal{L}}_{\mathcal{D}_{n}}(\boldsymbol{\psi}_{\text{TL}})$ over $N$ radar operating environments \;
			update shared parameter vector $\boldsymbol{\psi}_{\text{TL}}$ via (\ref{eq: transfer sgd})
		}
		\BlankLine
		\tcc{adaptation stage}
		initialize $\boldsymbol{\phi}^{(m)}=\boldsymbol{\psi}_{\text{TL}}$, and set $m=0$\;
		\While{stopping criterion not satisfied}{
			update receiver parameter vector $\boldsymbol{\phi}$ via (\ref{eq: rx sgd})
		}
		\caption{Design of Fast Adaptive Detector via Transfer Learning}
	\end{algorithm}
	\vspace{-2ex}
	
	\subsection{Design of Fast Adaptive Detector via Meta-Learning}
	As discussed in Section I, MAML aims to find the inferred inductive bias in the form of the initialization of the receiver neural network, denoted as $\boldsymbol{\psi}_{\text{MAML}}$, to enable fast adaptation on the current operating environment. 
	In the offline training stage, 
	the dataset $\mathcal{D}_n$ for the $n$th operating environment is randomly divided into two subsets. One subset of $\mathcal{D}_n$ is referred to as \emph{support set} $\mathcal{D}_n^s$, which is used to update the local receiver parameter $\boldsymbol{\theta}_n$ for the $n$th operating environment. The other subset of $\mathcal{D}_n$ is referred to as \emph{query set} $\mathcal{D}_n^q$, which is used to estimate the meta-learning empirical loss. Mathematically, with a single SGD iteration, we obtain the local update $\boldsymbol{\theta}_n$ according to the support set $\mathcal{D}_n^s$ for the $n$th operating environment  
	\begin{equation}
		\boldsymbol{\theta}_{n}=\boldsymbol{\psi}_{\text{MAML}}-\alpha \nabla _{\boldsymbol{\psi }_{\text{MAML}}
		}{\mathcal{L}}_{\mathcal{D}_{n}^{s}}(\boldsymbol{\psi}_{\text{MAML}}). \label{eq: local_update}
	\end{equation}%
	
	Based on the query set $\mathcal{D}_n^q$ and the local update $ \boldsymbol{\theta}_n$ (\ref{eq: local_update}), the meta-training
	empirical loss is given by 
	\begin{equation}
		\begin{aligned}
			{\mathcal{L}}(\boldsymbol{\psi }_{\text{MAML}})&=\sum_{n=1}^{N_b}{\mathcal{L}}_{\mathcal{D}_{n}^{q}}(%
			\boldsymbol{\theta}_{n})\\
			&=\sum_{n=1}^{N_b}{\mathcal{L}}_{\mathcal{D}_{n}^{q}}\big(
			\boldsymbol{\psi}_{\text{MAML}}-\alpha \nabla _{\boldsymbol{\psi }_{\text{MAML}}
			}{\mathcal{L}}_{\mathcal{D}_{n}^{s}}(\boldsymbol{\psi}_{\text{MAML}})\big),
		\end{aligned}
		\label{eq:empirical}
	\end{equation}
	where $N_b$ represents the number of environments selected randomly from $N$ offline operating environments for each meta training update, and is referred to as \emph{meta batch size}.
	The initialization of the receiver parameter vector $\boldsymbol{\psi}_{\text{MAML}}$ is learned via minimizing (\ref{eq:empirical}) through SGD
	\begin{equation}
		\begin{aligned}
			\boldsymbol{\psi}_{\text{MAML}}\leftarrow \boldsymbol{\psi}_{\text{MAML}}-\beta \nabla _{\boldsymbol{\psi}_{\text{MAML}}}{\mathcal{L}}(\boldsymbol{\psi}_{\text{MAML}})  &= \boldsymbol{\psi}_{\text{MAML}} - \beta\sum_{n=1}^{N_b} \nabla _{\boldsymbol{\psi}_{\text{MAML}}}{\mathcal{L}}_{\mathcal{D}_{n}^{q}}(\boldsymbol{\theta}_{n})\\
			& = \boldsymbol{\psi}_{\text{MAML}} - \beta \sum_{n=1}^{N_b}\big(\boldsymbol{I}-\alpha \nabla _{\boldsymbol{\psi}_{\text{MAML}}}^2{\mathcal{L}}_{\mathcal{D}_{n}^{s}}(\boldsymbol{\psi}_{\text{MAML}})  \big) \nabla_{\boldsymbol{\theta}_n}\mathcal{L}_{\mathcal{D}_{n}^{q}}(\boldsymbol{\theta}_{n}).
		\end{aligned}
		\label{eq:initial_update} 
	\end{equation}
	Note that the update (\ref{eq:initial_update}) requires calculating the second-order gradient $\nabla _{\boldsymbol{\psi}_{\text{MAML}}}^2{\mathcal{L}}_{\mathcal{D}_{n}^{s}}(\boldsymbol{\psi}_{\text{MAML}})$, which may be treated as constant and further ignored to speed up offline training time \cite{MAML2017}.
	
	During the adaptation stage, in a manner similar to (\ref{eq: rx sgd}), the receiver parameter vector $\boldsymbol{\phi}$ is updated according to the gradient ${\nabla}_{\boldsymbol{\phi}}{\mathcal{L}}_{\mathcal{D}_{a}}(\boldsymbol{\phi})$ with ${\boldsymbol{\phi}}^{(0)}=\boldsymbol{\psi}_{\text{MAML}}$. The algorithm of the design of fast adaptive detector via MAML is summarized in Algorithm 2.	
	\begin{algorithm}
		\DontPrintSemicolon
		\SetAlgoLined	
		\tcc{offline training stage}
		initialize parameter vector $\boldsymbol{\psi}_{\text{MAML}}$\;
		\While{stopping criterion not satisfied}{
			select $N_b$ environments randomly from $N$ offline operating environments\;
			\For{each selected operating environment $n$}{
				compute local parameter vector $\boldsymbol{\theta}_n$ from 
				(\ref{eq: local_update}) based on $\mathcal{D}_n^{s}$
			}
			compute meta-training empirical loss ${\mathcal{L}}(\boldsymbol{\psi}_{\text{MAML}})$ from (\ref{eq:empirical}) based on $\mathcal{D}_n^{q}$\;
			update parameter vector $\boldsymbol{\psi}_{\text{MAML}}$ via (\ref{eq:initial_update})
		}
		\BlankLine
		\tcc{adaptation stage}
		initialize $\boldsymbol{\phi}^{(m)}=\boldsymbol{\psi}_{\text{MAML}}$, and set $m=0$\;
		\While{stopping criterion not satisfied}{
			update receiver parameter vector $\boldsymbol{\phi}$ via (\ref{eq: rx sgd})
		}
		\caption{Design of Fast Adaptive Detector via Meta-Learning}
	\end{algorithm}

	\section{Numerical Results}
	This section first introduces models and parameters used in the simulation setup, and provides numerical results to evaluate the detection performance of the two proposed approaches.
	
	\subsection{Models and Parameters}
	The target is assumed stationary with a Rayleigh envelope $\alpha\sim\mathcal{CN}(0, \sigma_{\alpha}^2)$, where $\sigma_{\alpha}^2$ is the target power. The noise vector $\mathbf{n}$ has a zero-mean, complex Gaussian distribution with correlation matrix $\boldsymbol{\Omega}_n = \sigma^2_w \mathbf{I} + \sigma_I^2\boldsymbol{\Omega}_I$, where $\sigma^2_w$ is the thermal noise power level, $\sigma^2_I$ is signal-independent interference power level, and $\boldsymbol{\Omega}_I$ is the correlation matrix of the signal-independent interference. The signal-to-noise ratio is defined as $\text{SNR}\triangleq 10\log_{10}\{\sigma_{\alpha}^2/\sigma_w^2\}$. The signal-to-interference ratio is defined as $\text{SIR}\triangleq 10\log_{10}\{\sigma_{\alpha}^2/\sigma_I^2\}$. 
	The signal-independent interference is located in the frequency band $[f_l, f_u]$, where $f_l$ and $f_u$ represent the lower and upper frequencies normalized by the sampling frequency $f_s$, respectively. Accordingly, the correlation matrix is $[\boldsymbol{\Omega}_I]_{v,h}=f_u-f_l$ if $v=h$, and $[\boldsymbol{\Omega}_I]_{v,h}=[e^{j2\pi f_u (v-h)}-e^{j2\pi f_l (v-h)}]/[j2\pi(v-h)]$ otherwise, with $(v,h)\in\{1,\ldots,K\}^2$.
	The clutter vector $\mathbf{c}$ is the superposition of returns from $2K-1$  range cells \cite{Stoica2012}, namely
	\begin{equation}
		{\mathbf{c}}=\sum_{\substack{ g=-K+1 }}^{K-1}{\gamma }_{g}\mathbf{J%
		}_{g}{\mathbf{y}},
	\end{equation}
	where $\mathbf{J}_g$ and $\gamma_g$ represent the shift matrix and the complex clutter scattering coefficient for the $g$th range cell, respectively. Elements of the shift matrix is given by $[\mathbf{J}_g]_{v,h}=1$ if $v-h=g$, and $[\mathbf{J}_g]_{v,h}=0$ otherwise, with $(v,h)\in\{1,\ldots,K\}^2$. The clutter scattering coefficient $\gamma_g$ follows coherent Weibull distribution
	with shape parameter $\lambda$ and median ${\sigma}_m$ \cite{Richards2010}. Note that the nominal range of the shape parameter is $0.25\leq\lambda\leq2$ \cite{shape}.
	When the shape parameter $\lambda=2$, the clutter scattering coefficients are complex Gaussian random variables. Based on the assumed mathematical models of target, clutter, and noise, the optimal detector in the NP sense is available (see Appendix D of \cite{Wei2021} for details).
	
	The coded waveform $\mathbf{y}$ is a unit norm linear frequency modulated pulse with $K=16$ chips, namely $\mathbf{y} (k)= e^{j\pi R(k/f_s)^2} / \sqrt{K}$,
	for $k=0,\ldots, K-1$ with a chip rate $R=(100\times10^3)/(40\times10^{-6})$ Hz/s and a sampling rate $f_s=200$ kHz. A homogeneous clutter environment is considered. The median of the clutter scattering coefficient is set to $\sigma_m=0.0004$. The offline training stage is performed at $\text{SNR}_{\text{tr}}= 24\text{ dB}$. The offline training stage consists of $N=40$ different operating environments, which include two types of clutter distributions $\lambda_{\text{tr}}\in\{0.25, 2\}$, two signal-to-interference ratios $\text{SIR}_{\text{tr}}\in \{10\text{ dB}, 17\text{ dB}\}$, and ten different correlation matrices $\boldsymbol{\Omega}_I$ with a fixed frequency difference between the upper and lower normalized frequencies, i.e., $f_{u, \text{tr}}-f_{l,\text{tr}}=0.1$. The offline dataset for each operating environment $\mathcal{D}_n$ consists of $Q=4\times 10^5$ samples, equally divided between the $\mathcal{H}_0$ an $\mathcal{H}_1 $ hypotheses. 
	
	The adaptation stage is performed at $\text{SNR}_{\text{a}}=20\text{ dB}$ and $\text{SIR}_{\text{a}}=16\text{ dB}$. The upper and lower normalized frequencies are $f_{u, a}=0.6$ and $f_{l, a}=0.4$, respectively. The adaptation dataset $\mathcal{D}_a$ contains $Q_a=8000$ samples, equally divided between $\mathcal{H}_0$ and $\mathcal{H}_1$ hypotheses. Unless stated otherwise, we choose $40$ gradient updates to do adaptation for the two proposed approaches, i.e., $m=40$.
	
	In the testing stage, $2\times 10^5$ samples under hypothesis $\mathcal{H}_0$ are used to estimate the probability of false alarm $P_{fa}$, while $5\times10^4$ samples under hypothesis $\mathcal{H}_1$ are used to estimate the probability of detection $P_{d}$. Unless stated otherwise, the testing stage is performed at $\text{SNR}_{\text{te}}=13\text{ dB}$. 
	Receiver operating characteristic (ROC) curves are obtained via Monte Carlo simulation by varying the threshold applied at the output of the receiver.

	The receiver is a feedforward neural network with four layers, i.e., an input layer with $2K=32$ neurons, two hidden layers with $48$ neurons, and an output layer with $1$ neuron. The sigmoid function is adopted as the activation function for all neurons. In the offline training stage, we use a minibatch of size 128. The learning rates are set to $\alpha=0.2$ and $\beta=0.002$, respectively. For the design of fast adaptive detector via MAML, the meta batch size is set to $N_b=10$. In the adaptation stage, the batch of  adaptation samples is adopted to estimate the empirical cross-entropy loss (\ref{eq: rx loss grad}) with the learning rate $\alpha=0.002$.

	\subsection{Results}
	
\begin{figure}[H]
		\vspace{-4ex} \hspace{20ex} \includegraphics[width=0.6
		\linewidth]{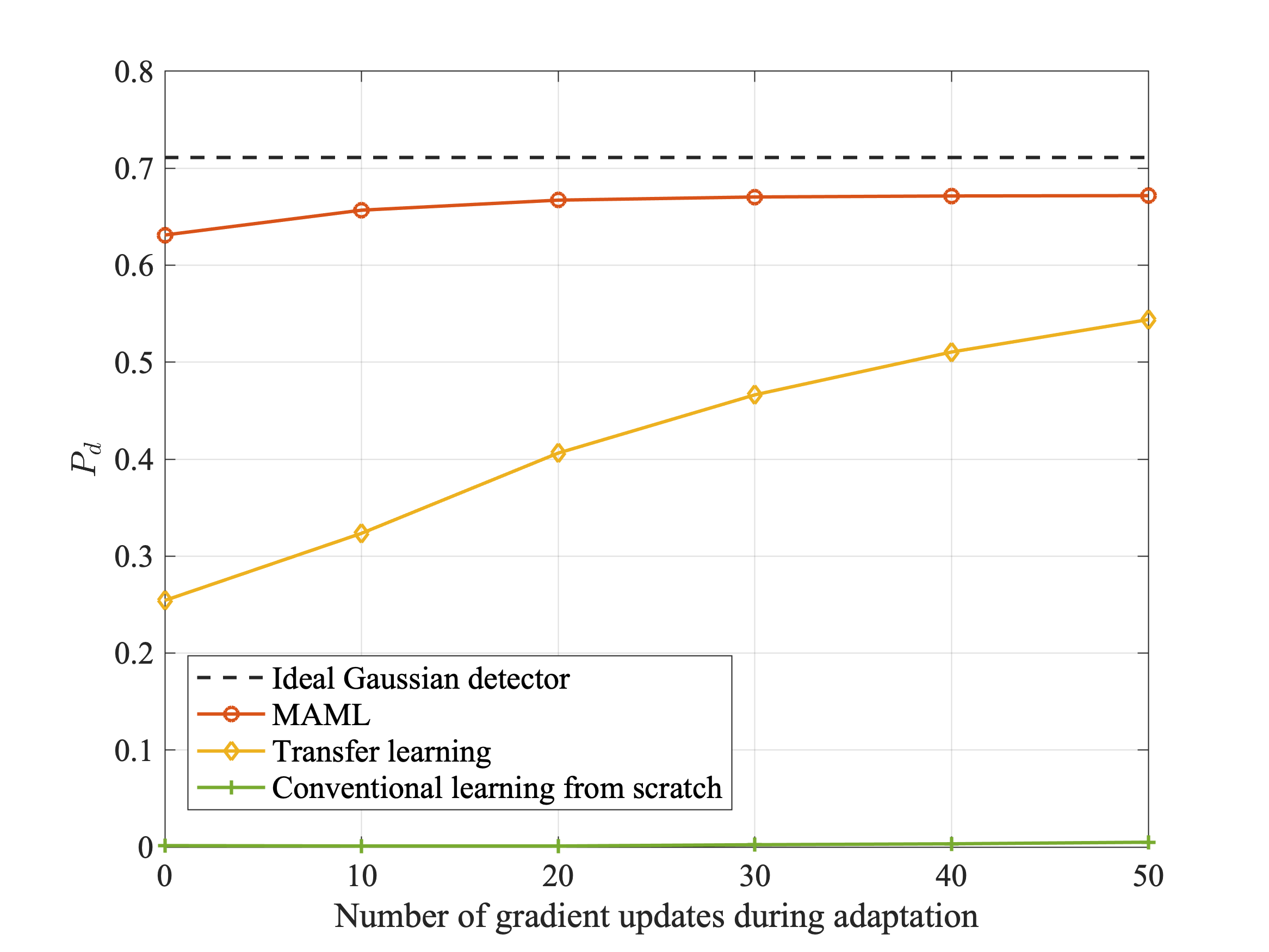} \vspace{-2ex}
		\caption{Illustration of the adaptation capability of the two proposed approaches in the presence of Gaussian clutter with $P_{fa}=10^{-3}$.}
		\label{f:adapt_2}
	\end{figure}

	Fig. \ref{f:adapt_2} illustrates the adaptation capability of the two proposed approaches in the presence of Gaussian clutter, i.e., $\lambda=2$, given the probability of false alarm $P_{fa} = 10^{-3}$. We compare the performance of the two proposed approaches with: (1) training from scratch (no prior knowledge) \cite{Moya2013}, whereby offline training data $\mathcal{D}$ is not used, and the receiver parameter vector $\boldsymbol{\phi}$ is initialized randomly; (2) an ideal Gaussian detector \cite{Wei2021} having access to the actual operating environment conditions. As shown in the figure, the detection performance of the proposed two approaches is close to the upper bound set by the ideal Gaussian detector  \cite{Wei2021} even for a small number of gradient updates. Moreover, MAML is seen to adapt faster as compared with transfer learning. In contrast, conventional learning from scratch performs poorly due to the limited number of gradient updates. 

\begin{figure}[H]
		\vspace{-2ex} \hspace{20ex} \includegraphics[width=0.6
		\linewidth]{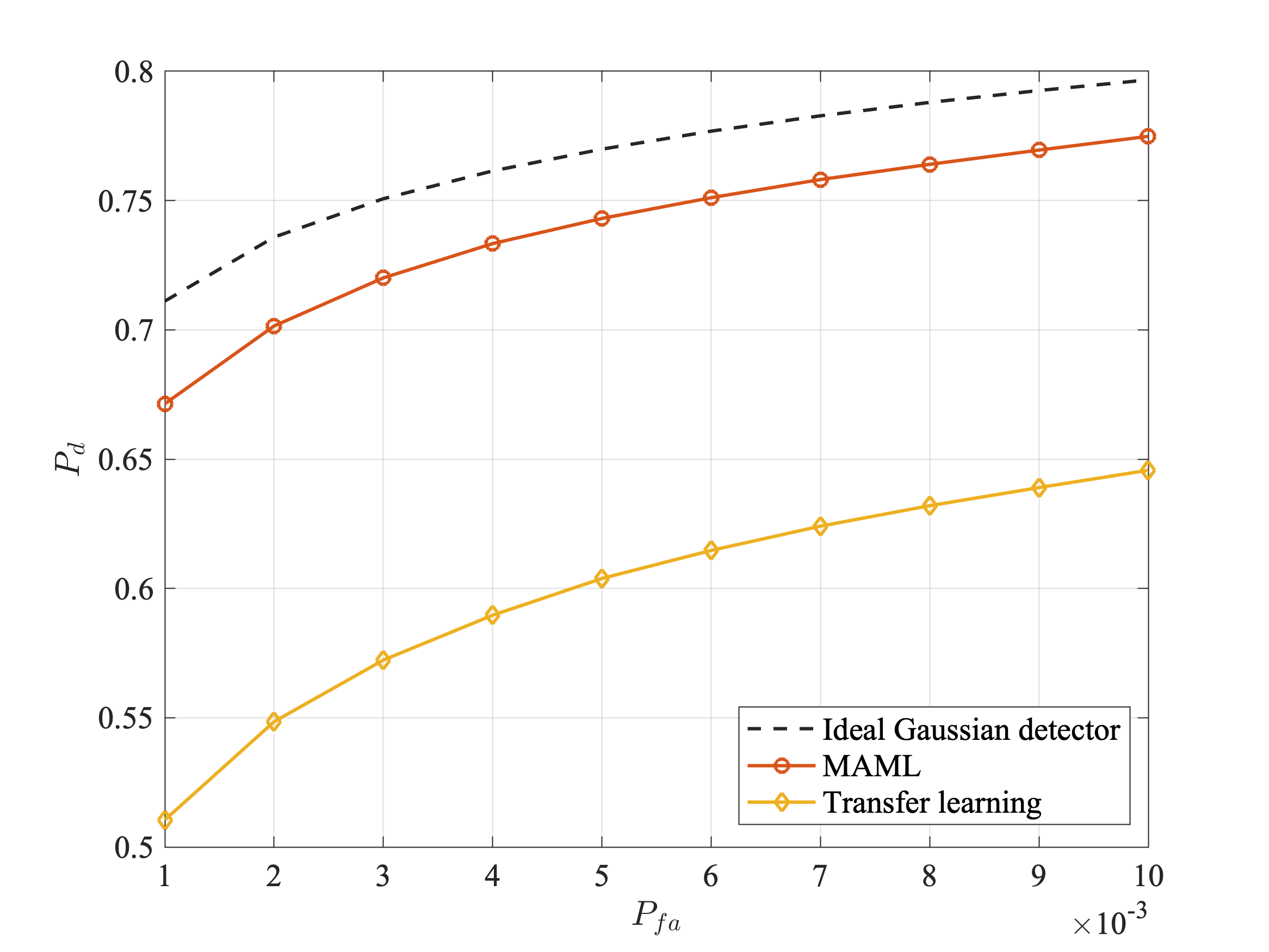} \vspace{-2ex}
		\caption{ROC curves of transfer learning-based detector and MAML-based detector in Gaussian clutter.}
		\label{f:roc_2}
	\end{figure}

	Fig. \ref{f:roc_2} compares ROC curves of the two proposed approaches in Gaussian clutter with the fixed number of gradient updates during the adaptation. With the limited number of gradient updates, the ROC curve obtained based on conventional learning from scratch is not shown due to poor performance. It is observed that the MAML-based detector outperforms the transfer learning-based detector in the presence of Gaussian clutter. For instance, for $P_{fa}=5\times10^{-3}$, MAML-based detector yields $P_d=0.74$, while transfer learning-based detector yields $P_d=0.6$.
	
\begin{figure}[H]
		\vspace{-2ex} \hspace{20ex} \includegraphics[width=0.6
		\linewidth]{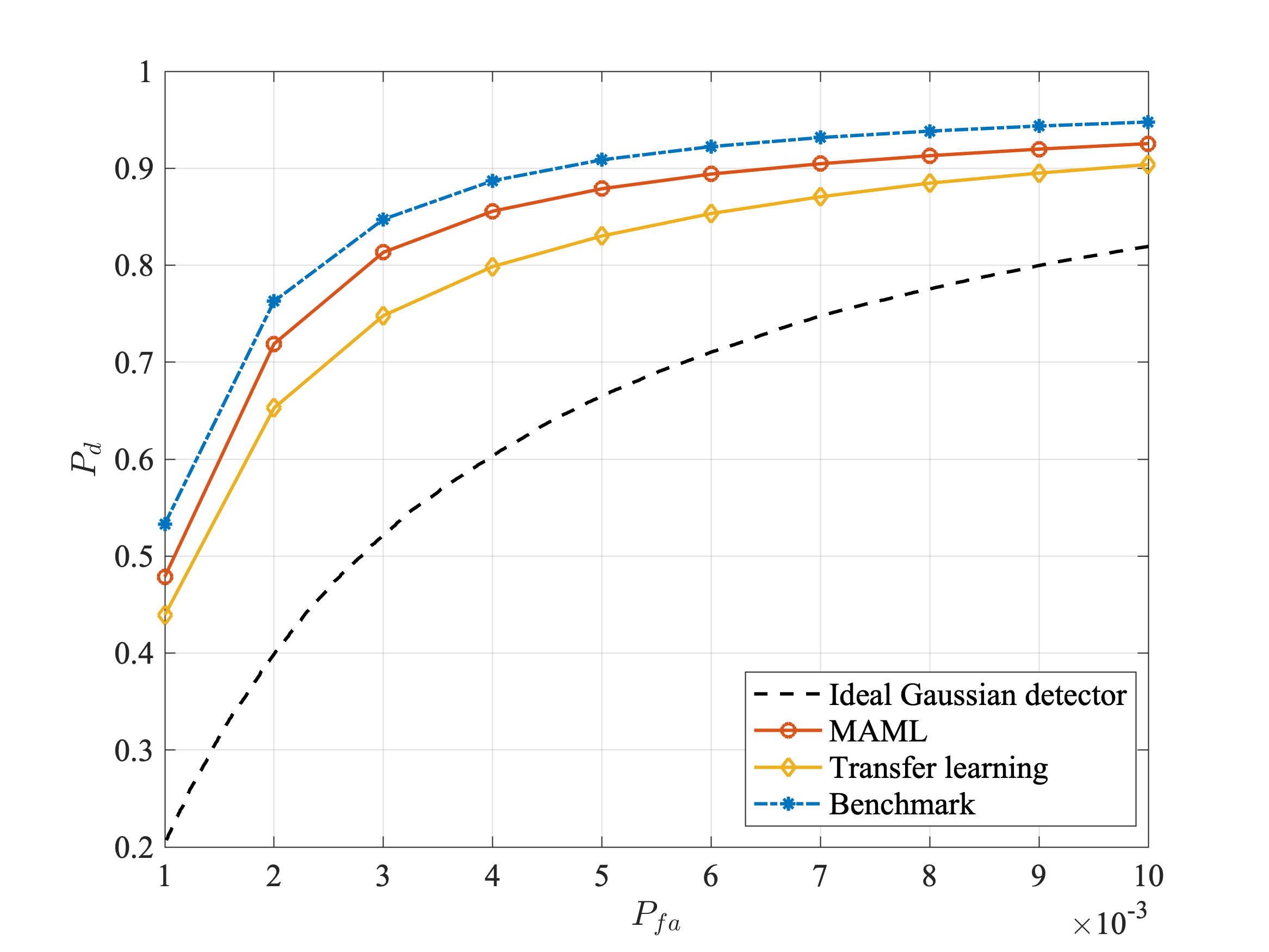} \vspace{-2ex}
		\caption{ROC curves of transfer learning-based detector and MAML-based detector in non-Gaussian clutter.}
		\label{f:roc_025}
	\end{figure}

	Detection performance of the proposed two approaches in non-Gaussian clutter $\lambda=0.25$ is shown in Fig. \ref{f:roc_025}. When the clutter is non-Gaussian, the optimal detector is not available. Thus, receiver training with a large number of training samples and gradient updates is adopted as the benchmark. The testing stage is performed at $\text{SNR}_{\text{te}}=25$ dB. As shown in the figure, both the transfer learning-based detector and MAML-based detector provide comparable detection performance with the benchmark. Moreover, the ideal Gaussian detector  \cite{Wei2021} provides the worst detection performance due to the mismatch between the assumed mathmatical models  and the actual non-Gaussian testing environment. 
	
	\section{Conclusions}
	This paper proposes two methods of learning radar detectors. Each method
	comprises an offline training stage and an adaptation stage. We have developed two offline training algorithms, both of which enable fast adaptation of detectors with limited data. The offline training stage may be implemented either via transfer learning or meta-learning. Numerical results have shown that the proposed two approaches make remarkable gains over the receiver training with no prior knowledge. Moreover, the meta-learning-based detector outperforms the transfer learning-based detector in both Gaussian and non-Gaussian clutter.

	\section*{Acknowledgment}
	Research was sponsored by the Army Research Laboratory and was accomplished under
	Cooperative Agreement Number W911NF-20-2-0219. The views and conclusions contained in this document are
	those of the authors and should not be interpreted as representing the official policies, either expressed or implied, of
	the U.S. Government. The U.S. Government is authorized to reproduce and
	distribute reprints for Government purposes notwithstanding any copyright notation herein.

\end{document}